\documentclass{article}
\usepackage{spconf,amsmath,graphicx,hyperref}
\usepackage{multirow}
\usepackage{float} 
\usepackage{titlesec}
\usepackage{balance}
\usepackage{amssymb,amsfonts}
\usepackage{algorithm}
\usepackage{algorithmic}
\usepackage{booktabs}
\usepackage{graphicx}
\usepackage{array}
\usepackage{xcolor}
\usepackage{url}
\usepackage[T1]{fontenc}

\newcommand{\x}{\mathbf{x}}
\newcommand{\y}{\mathbf{y}}

\newcommand{\sg}{\operatorname{sg}}
\newcommand{\Ltext}[1]{\mathcal{L}_{\mathrm{#1}}}

\def\x{{\mathbf x}}

\title{Teacher-Free Self-Distilled Consistency Trajectory\\ Learning for Fast Speech Enhancement}
\name{Shuubham Ojha and Carol Espy-Wilson}
\address{
    Institute of Systems Research, Dept. of Electrical and Computer Engineering \\
    University of Maryland,  College Park, MD, USA 
}  
\begin{document}
%
\maketitle
\begin{abstract}
Consistency trajectory models offer a route to fast, high-quality speech
enhancement, collapsing the many reverse steps of diffusion-based enhancers into
a handful. When instantiated on a Schr\"odinger bridge (SB), which pins the
generative process to fixed clean and noisy endpoints, existing
consistency-trajectory enhancers (SBCTMs) still require a pretrained
teacher to supply trajectory supervision, which raises training cost and ties the
final quality to that of the teacher. We propose a teacher-free, self-distilled
consistency-trajectory framework that removes the external teacher resulting in a $5$X reduction in per epoch training time. Our model is trained with a three-stage curriculum of clean speech prediction, a
self-distilled shortcut objective, and perceptual fine-tuning with a
multi-resolution short-time Fourier transform (MR-STFT) loss. Using the same
NCSN++ backbone as SBCTM,
our model attains a wide-band PESQ of $3.01$, ESTOI $0.87$, and SI-SDR
$19.07$\,dB on VoiceBank+DEMAND compared to $3.57$, $0.87$ and $12.8$\,dB for the teacher based model. Further, we find that a geometric
schedule at low reverse step count maximizes perceptual quality, while a higher-step
uniform schedule favors signal fidelity.
\end{abstract}


\section{Introduction}
Casting speech enhancement (SE) as a generative process with score-based and
diffusion models has achieved state-of-the-art perceptual quality on standard
benchmarks \cite{welker2022sgmse,richter2023sgmse,lemercier2023storm}. A recurring
obstacle is inference latency, since such enhancers typically require tens to
hundreds of reverse denoising steps to produce a
clean-speech estimate.
 
Consistency models \cite{song2023consistency}, consistency trajectory models
(CTMs) \cite{kim2024ctm}, and consistency flow matching \cite{yang2024cfm}
accelerate this reverse denoising process by learning a trajectory (or velocity) function that jumps between
points along the generative path, enabling few-step generation while preserving a
favorable quality and speed trade-off. The Schr\"odinger bridge (SB) suits SE
because, unlike unconditional diffusion, it defines a process with fixed endpoints
(clean speech at one end, the noisy observation at the other), removing the prior
mismatch between the end of the forward process and the start of the reverse
process \cite{jukic2024sb,debortoli2021dsb}. Combining the two, SB consistency
trajectory models (SBCTMs) \cite{nishigori2025sbctm} give large real-time-factor
gains over diffusion SB baselines.
 
A key limitation of SBCTM, inherited from CTM, is its reliance on a separately
pretrained teacher, since trajectory targets are generated by the teacher and the
student is distilled against them. Such a model requires the teacher's inference at
each training step, raising the training and memory cost, and caps the attainable
student quality at that of the teacher. This raises the question if
consistency-trajectory learning for SE can be achieved without a pretrained teacher.
Drawing on consistency training \cite{song2023consistency} and mean-teacher
self-distillation \cite{tarvainen2017mean}, we replace the external teacher with
an Exponential Moving Average (EMA) copy of the student, which bootstraps trajectory targets during training.
 
Our contributions are: (i)~a teacher-free consistency-trajectory framework in
which an EMA self-distillation target replaces the pretrained teacher, removing a
full training run; (ii)~a three-stage curriculum ($\x_0$ prediction, a
self-distilled shortcut objective, perceptual fine-tuning) on the
variance-exploding SB marginal; (iii)~with the backbone held identical to SBCTM, a
model with two-step inference reaching PESQ $3.01$, ESTOI $0.87$, and SI-SDR $19.07$\,dB on
VoiceBank+DEMAND; and (iv)~a step-count and
schedule sweep that reveals a perception-fidelity trade-off, with a geometric
schedule at low step counts maximizing perceptual quality and a higher-step
uniform schedule favoring signal fidelity.

\section{Related Work}
Score-based and diffusion models formulate
enhancement as iterative denoising in the complex STFT domain
\cite{welker2022sgmse,richter2023sgmse, lemercier2023storm}. They reach excellent quality but need many reverse iterations. On the other hand, CTM 
\cite{song2023consistency, kim2024ctm} learns a function that maps any point on a
probability-flow Ordinary Differential Equation (ODE) trajectory to that trajectory's shared endpoint; it can be
trained by distillation from a pretrained score model, or without one by enforcing
this self-consistency directly along the trajectory. SBCTM \cite{nishigori2025sbctm} applies CTM
learning to the SB with a differentiable-PESQ auxiliary loss, but its student is
distilled against targets produced by a separately pretrained teacher at
inference. We remove that dependence, learning the trajectory function teacher-free
via EMA self-distillation and using an MR-STFT perceptual objective in place of
the PESQ loss.

\begin{figure}[t]
  \centering
  \includegraphics[width=\columnwidth]{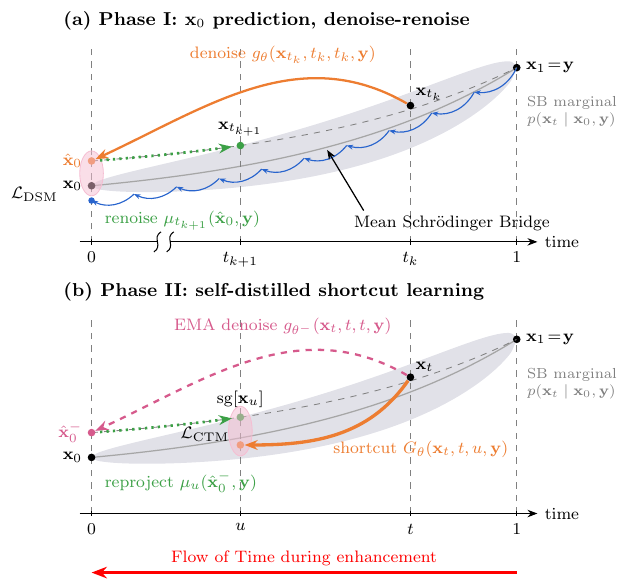}
  \caption{Training overview of our model (a) Phase~I trains the boundary denoiser $g_\theta$ by $\mathbf{x}_0$ regression by denoising-renoising. (b) Phase~II forms the teacher-free target from Phase~I, and the student learns intermediate jumps. The dashed curve is an estimate of the mean bridge, re-anchored at the clean-speech estimate $\hat{\mathbf{x}}_0$, rather than the clean speech ${\mathbf{x}}_0$ itself. Small blue steps illustrate a conventional many-step model (eg. diffusion).}
  \label{fig:overview}
\end{figure}

As commonplace in SE literature, we denote the clean speech by $\x_0$ and the noisy observation by $\y$ (Fig.~\ref{fig:overview}), both as complex
STFT coefficients \cite{richter2023sgmse}. The SB defines a process
$\{\x_t\}_{t\in[0,1]}$ pinned to $\x_0$ (clean, $t{=}0$) and $\y$ (noisy,
$t{=}1$). We adopt the variance-exploding SB marginal of Juki\'c et al.\
\cite{jukic2024sb}, as used in \cite{nishigori2025sbctm}. With base $k$ and
diffusion $c$,
\begin{equation}
  \sigma_t = \sqrt{\tfrac{c\,(k^{2t}-1)}{2\ln k}}, \quad
  \sigma_1 = \sigma_t|_{t=1}, \quad
  \bar{\sigma}_t = \sqrt{\sigma_1^2 - \sigma_t^2}.
  \label{eq:sbve-sigmas}
\end{equation}
The conditional marginal is
$p(\x_t\mid\x_0,\y)=\mathcal{N}(\x_t;\boldsymbol{\mu}_t,\varsigma_t^2\mathbf{I})$, with
\begin{equation}
  \boldsymbol{\mu}_t = w_x(t)\,\x_0 + w_y(t)\,\y, \qquad
  \varsigma_t = \tfrac{\bar{\sigma}_t\,\sigma_t}{\sigma_1},
  \label{eq:sbve-mean}
\end{equation}
where $w_x(t)=\bar{\sigma}_t^2/\sigma_1^2$ and $w_y(t)=\sigma_t^2/\sigma_1^2$. At
$t{=}0$, $\sigma_0{=}0$ gives $w_x{=}1$, $w_y{=}0$, $\varsigma_0{=}0$, so $\x_0$ is
recovered; at $t{=}1$, $\bar{\sigma}_1{=}0$ gives $w_x{=}0$, $w_y{=}1$, so
$\x_1{=}\y$. We use $k{=}2.6$, $c{=}0.05$. Enhancement transports $\x_1{=}\y$ back to
$t{=}0$. While diffusion samplers discretize this into many steps,
consistency-trajectory models use shortcuts to avoid multiple passes.

\section{Proposed Method}
\subsection{Trajectory parameterization}
Let $g_\theta(\x_t,t,s,\y)$ be a network conditioned on the noisy observation
$\y$, a source time $t$, and a target time $s\le t$, with output on the scale of
clean speech. Following the CTM parameterization \cite{kim2024ctm},
\begin{equation}
  G_\theta(\x_t,t,s,\y) = \tfrac{s}{t}\,\x_t
  + \big(1-\tfrac{s}{t}\big)\, g_\theta(\x_t,t,s,\y).
  \label{eq:gparam}
\end{equation}
The clean-speech estimate is the raw network at $s{=}t$, that is
$\hat{\x}_0 = g_\theta(\x_t,t,t,\y)$, while the skip term carries the input state at training time,
when the target time equals the source time. During inference, we proceed on a decreasing time grid $1 = t_N > t_{N-1} > \cdots > t_0$ from $\x_{t_N} = \y$. 
 
\subsection{Phase I: clean-speech prediction}
We warm-start the denoiser by regression,
\begin{equation}
  \Ltext{DSM}(\theta)
  = \mathbb{E}\big[\,| g_\theta(\x_t,t,t,\y) - \x_0 |^2 \,\big],
  \label{eq:dsm}
\end{equation}
with $\x_t\sim p(\x_t\mid\x_0,\y)$ and $t\sim\mathcal{U}[t_{\min},1]$,
$t_{\min}{=}0.03$. This term is retained as an anchor later. Since the Phase-I model is trained
only at the boundary $s{=}t$ and never learns the shortcut
$G_\theta(\cdot, t, s, \cdot)$ for $s < t$, so the shortcut-grid sampler cannot be
applied to it. We instead use a denoise-renoise scheme (Fig.~\ref{fig:overview}(a)) that relies solely on the
boundary denoiser and the analytic SB marginal \eqref{eq:sbve-mean}. Starting from
$\x = \y$, at each grid time $t_i$ we (i)~\emph{denoise}, predicting the clean
estimate $\hat{\x}_0 = g_\theta(\x, t_i, t_i, \y)$, and (ii)~\emph{renoise},
reprojecting that estimate onto the bridge at the next, lower time via the marginal
mean, $\x \leftarrow \boldsymbol{\mu}_{t_{i+1}}(\hat{\x}_0, \y)$. Thus, the final grid
point returns $\hat{\x}_0$. 

 
\subsection{Phase II: self-distilled shortcut learning}
Rather than a pretrained teacher, we keep an EMA copy of the student,
$\theta^- \leftarrow \mu\theta^- + (1-\mu)\theta$, with decay $\mu{=}0.999$. Times
are sampled on a Karras power \cite{karras2022edm} : for $N_{\max}$ levels and exponent $\rho$,
\begin{equation}
  \tau(i) = \Big(1 + \tfrac{i}{N_{\max}-1}\big(t_{\min}^{1/\rho}-1\big)\Big)^{\rho},
  \label{eq:karras}
\end{equation}
from which a source index and a positive step gap yield $t=\tau(i)$,
$u=\tau(i{+}\Delta)$ with $u<t$ ($\rho{=}7$, $N_{\max}{=}40$). Given
$\x_t\sim p(\x_t\mid\x_0,\y)$, the teacher-free target is one EMA denoising step
followed by a deterministic reprojection onto the bridge at $u$ via the SB mean
\eqref{eq:sbve-mean}:
\begin{equation}
  \hat{\x}_0^{-} = g_{\theta^-}(\x_t,t,t,\y), \quad
  \x_u = \boldsymbol{\mu}_u(\hat{\x}_0^{-},\y).
  \label{eq:teacherfree-target}
\end{equation}
The student learns the direct shortcut $t\!\to\!u$,
\begin{equation}
  \Ltext{CTM}(\theta)
  = \mathbb{E}\big[\,| G_\theta(\x_t,t,u,\y) - \sg[\x_u] |^2 \,\big],
  \label{eq:ctm}
\end{equation}
with stop-gradient $\sg[\cdot]$ as shown in Fig.~\ref{fig:overview}(b). As $\x_u$ comes from the model's own EMA, no
external teacher is needed: the reprojection plays the role of the teacher's
solver step in SBCTM. With a concurrent DSM anchor \eqref{eq:dsm} at an
independent time, the Phase-II objective is
$\mathcal{L} = \Ltext{CTM} + \lambda_{\mathrm{DSM}}\Ltext{DSM}$,
$\lambda_{\mathrm{DSM}}{=}1$.  Once the trajectory
shortcuts are trained ($t \rightarrow u$), we chain them directly as
$\x_{t_{k+1}} = G_\theta(\x_{t_k}, t_k, t_{k+1}, \y)$ during inference, and return $\x_{t_N}$ with each
step being a single network evaluation.
 
\subsection{Phase III: perceptual fine-tuning}
We add a waveform-domain MR-STFT loss on the ISTFT-reconstructed estimate over
three resolutions with FFT sizes $\{512,1024,2048\}$ and proportional hop and
window lengths. Each resolution contributes spectral-convergence and log-magnitude $\ell_1$ terms \cite{yamamoto2020pwg}.
The full objective is
\begin{equation}
  \mathcal{L}
  = \Ltext{CTM}
  + \lambda_{\mathrm{DSM}}\Ltext{DSM}
  + \lambda_{\mathrm{PER}}\Ltext{MR\text{-}STFT},
  \label{eq:total}
\end{equation}
with $\lambda_{\mathrm{DSM}}{=}1$, $\lambda_{\mathrm{PER}}{=}0.05$. Unlike SBCTM,
whose auxiliary term is a differentiable PESQ loss, our perceptual objective does
not optimize the evaluation metric directly.
 
\subsection{Inference steps and schedule}
\label{sec:inference}
\label{sec:geometric}
At the low step counts we target, the inference grid is a significant design
choice, and Table~\ref{tab:sched} shows both schedule
and step count to visibly affect quality. All schedules terminate at $t{=}0.03$
rather than $0$ (a slight abuse of the notation $t_N{\ge}0$), and we write an
$N$-step schedule for a grid of $N$ jumps. We adopt a geometric grid, in which the
intermediate nodes are warped toward the clean end, over a uniform one where the nodes are equally spaced. The source and target times of each shortcut are drawn from the
Karras power grid of Eq.~\eqref{eq:karras}, a non-uniform, warped
discretization that concentrates its levels toward $t_{\min}$; the network
therefore only ever learns jumps whose endpoints lie on that warped grid. Consequently, our best-PESQ operating point is the two-step geometric schedule $\{\tau(0), \tau(14), \tau(39)\} = \{1.0, 0.344, 0.03\}$ of Eq.~\eqref{eq:karras}(Table~\ref{tab:sched}).

\begin{table}[t]
\centering
\caption{Performance on VB+DEMAND ($824$ utterances). Higher is better except Inference Steps. DNSMOS is P.808.}
\label{tab:main}
\setlength{\tabcolsep}{2pt}
\begin{tabular}{lccccc}
\toprule
Method & $\#$ Steps & PESQ & ESTOI & SI-SDR & DNSMOS \\
\midrule
Noisy                      & n/a & 1.96 & 0.79 & 8.4 & 3.08 \\
SGMSE+ \cite{richter2023sgmse}       & 60  & 2.86 & 0.86 & 17.50 & 3.52 \\
StoRM \cite{lemercier2023storm}      & 60  & 2.89 & 0.86 & 18.79 & 3.51 \\
SBCTM \cite{nishigori2025sbctm}      & 4   & 3.57 & 0.87 & 12.8 & 3.54 \\
SE-Bridge \cite{qiu2023sebridge}     & 1   & 2.97 & 0.87 & 19.9 & n/a  \\
SB-PESQ \cite{richter2024objectives} & 4   & 3.55 & 0.87 & 13.0 & 3.54 \\
\midrule
\textbf{Ours} (2-step geo.)          & 2   & 3.01 & 0.87 & 19.07 & 3.51 \\
\bottomrule
\end{tabular}
\end{table}

\section{Experimental Setup}
We evaluate on VoiceBank+DEMAND \cite{valentini2016voicebank} at 16\,kHz over all
$824$ test utterances. Inputs are complex spectrograms (Hann window,
$n_{\mathrm{FFT}}{=}510$, i.e.\ 256 bins, hop $128$) with an amplitude-compression
transform (exponent $0.5$). The backbone $g_\theta$ is the NCSN++ trajectory
network (ncsnpp-ctm\_v2) with $66.6$\,M parameters, the same backbone as
SBCTM \cite{nishigori2025sbctm}. We optimize with Adam (learning rate $10^{-4}$, $5000$-step
warmup, gradient clipping at norm $1.0$) using a batch size of $2$ with gradient
accumulation over $8$ steps, for an effective batch size of $16$ to match SBCTM.
Phases~I and~II are each trained for $200{,}000$ optimizer steps and Phase~III for
$390{,}000$. Remaining settings are EMA decay $\mu{=}0.999$, SB parameters
$k{=}2.6$, $c{=}0.05$, and shortcut-grid parameters $\rho{=}7$, $N_{\max}{=}40$,
$t_{\min}{=}0.03$. We report wide-band PESQ (ITU-T P.862.2 at 16\,kHz), ESTOI,
SI-SDR \cite{leroux2019sisdr}, DNSMOS P.808 \cite{reddy2021dnsmos} and DNSMOS P.835 \cite{reddy2022dnsmos}
(SIG/BAK/OVRL). Baselines include SGMSE+ \cite{richter2023sgmse}, StoRM
\cite{lemercier2023storm}, the Brownian-bridge-based enhancer \cite{qiu2023sebridge},
the teacher model for SBCTM (SB-PESQ) \cite{richter2024objectives}, and the
teacher-based SBCTM \cite{nishigori2025sbctm}.

\section{Results}
Table~\ref{tab:main} reports the main comparison, Table~\ref{tab:cost} contrasts
SBCTM and the proposed model, and Table~\ref{tab:sched} studies step count and
schedule. The SBCTM PESQ in Table~\ref{tab:main} is the score reported in the
original SBCTM paper \cite{nishigori2025sbctm}. Our reported model is the two-step geometric schedule, reaching PESQ $3.01$, ESTOI
$0.87$, and SI-SDR $19.07$\,dB on the same backbone
as SBCTM, showing that a pretrained teacher is not required for competitive
consistency-trajectory enhancement. Table~\ref{tab:sched} sweeps step count and
schedule and reveals a perception-fidelity trade-off \cite{blau2018perception}
rather than a single dominant setting. Perceptual quality is maximized at low step
counts with a geometric grid: PESQ peaks at $3.01$ for two-step geometric, above
both the single-step model ($2.93$) and the three- and four-step geometric grids
($2.96$ and $2.94$). Signal fidelity, by contrast, is
maximized by the four-step uniform schedule (SI-SDR $20.07$\,dB),
which trails on PESQ ($2.95$). The geometric advantage
is clearest on PESQ at the lowest step count and on SI-SDR and DNSMOS at step count 3, while narrowing or even inverting on fidelity
metrics as the grid is refined, consistent with the training-matched argument of
Section~\ref{sec:geometric}. We report the two-step geometric model as our primary
configuration because it gives the best perceptual quality at the lowest inference
cost, and note that a fidelity-oriented deployment may prefer four-step uniform. A
likely contributor to the PESQ gap with respect to SBCTM is that SBCTM optimizes a
differentiable-PESQ loss, training directly toward the metric on which it is
evaluated, whereas our model is trained with a more general MR-STFT perceptual
objective.

\begin{table}[t]
\centering
\caption{SBCTM vs.\ the proposed teacher-free model. Both use the identical
\texttt{ncsnpp-ctm\_v2} backbone.}
\label{tab:cost}
\begin{tabular}{lcc}
\toprule
Property & SBCTM \cite{nishigori2025sbctm} & \textbf{Ours} \\
\midrule
Pretrained teacher required & Yes & \textbf{No} \\
Perceptual objective        & PESQ loss & MR-STFT \\
Inference Steps               & \textbf{2} & \textbf{2} \\
PESQ                        & \textbf{3.54} & 3.01 \\
SI-SDR                      & 13.2 & \textbf{19.07} \\
\bottomrule
\end{tabular}
\end{table}
 
\begin{table}[t]
\centering
\caption{Step count and schedule sweep (proposed model). PESQ is maximized by the
2-step geometric schedule (our reported model); signal fidelity (SI-SDR) by 4-step uniform and
DNSMOS is maximized by 3-step geometric.}
\label{tab:sched}
\setlength{\tabcolsep}{4pt}
\resizebox{\columnwidth}{!}{%
\begin{tabular}{llcccccc}
\toprule
 & & & & & \multicolumn{3}{c}{DNSMOS (P.835)} \\
\cmidrule(l){6-8}
Steps & Schedule & PESQ & ESTOI & SI-SDR & SIG & BAK & OVRL \\
\midrule
1 & n/a       & 2.93 & 0.87 & 19.98 & 3.46 & 4.03 & 3.18 \\
2 & Geometric & \textbf{3.01} & 0.87 & 19.07 & 3.49 & 3.97 & 3.17 \\
2 & Uniform   & 2.89 & 0.87 & 19.73 & 3.49 & 4.00 & 3.19 \\
3 & Geometric & 2.96 & 0.87 & 19.52 & \textbf{3.62} & \textbf{4.15} & \textbf{3.39} \\
3 & Uniform   & 2.83 & 0.87 & 18.69 & 3.50 & 3.95 & 3.17 \\
4 & Geometric & 2.95 & 0.87 & 19.44 & 3.51 & 3.99 & 3.19 \\
4 & Uniform   & 2.95 & \textbf{0.87} & \textbf{20.07} & 3.51 & 4.06 & 3.24 \\
\bottomrule
\end{tabular}%
}
\end{table}

\section{Ablation Study}
Table~\ref{tab:abl} isolates the contribution of each training stage at our
reported two-step geometric schedule. The Phase-I model has no trajectory shortcuts
and is evaluated with the denoise-renoise sampler of Section~\ref{sec:inference};
the later stages use the shortcut sampler. Phase~I alone is already a competent denoiser: it reaches PESQ $2.58$ and, notably,
the highest SI-SDR ($19.52$\,dB) of any configuration,
consistent with direct $\x_0$ regression favouring signal fidelity while leaving
perceptual quality lower. Adding the dedicated shortcut stage (Phase~II) is where
the perceptual gain is concentrated: PESQ rises to $2.99$ at a modest fidelity cost
(SI-SDR $19.28$), and perceptual fine-tuning (Phase~III) adds a small
further gain to PESQ $3.01$. The dedicated Phase-II stage is, however, not strictly required. Since the
shortcut (self-consistency) loss is itself part of the Phase-III objective
\eqref{eq:total}, a Phase~I~$\rightarrow$~Phase~III schedule that skips Phase~II
still learns the few-step map, reaching PESQ $2.95$ (within $0.06$ of the full
model) while retaining higher fidelity (SI-SDR $19.76$, OVRL $3.21$).

\begin{table}[t]
\centering
\caption{Curriculum ablation at the two-step geometric schedule. OVRL is the
DNSMOS P.835 overall score.}
\label{tab:abl}
\begin{tabular}{lcccc}
\toprule
Configuration & PESQ & ESTOI & SI-SDR & OVRL \\
\midrule
Phase I only                            & 2.58 & 0.85 & 19.52 & 3.18 \\
\ \ + Phase II                          & 2.99 & 0.87 & 19.28 & 3.19 \\
\ \ + Phase III (full)                  & \textbf{3.01} & 0.87 & 19.07 & 3.17 \\
\midrule
Phase I $\rightarrow$ III & 2.95 & 0.87 & 19.76 & 3.21 \\
\bottomrule
\end{tabular}
\end{table}

\section{Training Efficiency}
Our teacher-free formulation removes the need for a separately pretrained teacher model and permits training within a substantially smaller practical compute envelope. All experiments were performed on a single NVIDIA RTX~3090 with an effective batch size of 16. Our model uses a micro-batch of 2 with gradient accumulation over 8 steps and requires 19.3, 45.2, and 64.6 min per epoch for Phases~I, II, and~III, respectively. Peak allocated GPU memory is 6.1, 11.2, and 15.8 GB across the three phases. For comparison, SBCTM uses a micro-batch of 1 with 16 gradient-accumulation steps to obtain the same effective batch size of 16. Under this configuration, one training epoch requires 327.1 min (5.45 h), with 9.57 GB peak allocated memory. Increasing the SBCTM micro-batch to 2, thereby matching our micro-batch and accumulation configuration, results in an out-of-memory error on the same 24-GB RTX~3090. Thus, a strictly batch-matched wall-clock comparison is not possible on frugal hardware.

\section{Conclusion}
We presented a teacher-free, self-distilled consistency-trajectory framework for
speech enhancement that removes the pretrained teacher of SBCTM by using an EMA
copy of the student to generate trajectory targets on a variance-exploding
Schr\"odinger bridge. Trained with a three-stage curriculum and evaluated on the
same backbone as SBCTM, our two-step geometric model attains PESQ $3.01$, ESTOI
$0.87$, and SI-SDR $19.07$\,dB on VoiceBank+DEMAND.
A sweep over step count and schedule reveals a perception-fidelity trade-off: a
geometric grid at low step counts maximizes perceptual quality (PESQ), while a
higher-step uniform grid favors signal fidelity, with the geometric
advantage narrowing as the grid is refined.

\bibliographystyle{IEEEbib}
\bibliography{strings,refs}

\end{document}